\documentclass[twocolumn]{pasj00}

\draft

\usepackage{natbib}
\bibpunct{(}{)}{;}{a}{}{,}

\usepackage{lscape}
\usepackage{longtable}

\begin{document}
\SetRunningHead{S. Takano et al.}{Classification of molecular distributions
in NGC 1068 observed with ALMA}

\title{Distributions of molecules in the circumnuclear
disk and surrounding starburst ring in the Seyfert galaxy 
NGC 1068 observed with ALMA}

\author{
Shuro \textsc{Takano}\altaffilmark{1,2},
Taku \textsc{Nakajima}\altaffilmark{3},
Kotaro \textsc{Kohno}\altaffilmark{4,5},
Nanase \textsc{Harada}\altaffilmark{6},
Eric \textsc{Herbst}\altaffilmark{7},
Yoichi \textsc{Tamura}\altaffilmark{4},
Takuma \textsc{Izumi}\altaffilmark{4},
Akio \textsc{Taniguchi}\altaffilmark{4},
and
Tomoka \textsc{Tosaki}\altaffilmark{8}
}
\altaffiltext{1}{Nobeyama Radio Observatory,
Nobeyama, Minamimaki, Minamisaku,
Nagano 384-1305}
\email{takano.shuro@nao.ac.jp}
\altaffiltext{2}{Department of Astronomical Science,
The Graduate University for Advanced
  Studies (Sokendai),
Nobeyama, Minamimaki, Minamisaku,
Nagano 384-1305}
\altaffiltext{3}{The Solar-Terrestrial Environment 
Laboratory, Nagoya University,
Furo-cho, Chikusa-ku, 
Nagoya 464-8601}
\altaffiltext{4}{Institute of Astronomy, University of Tokyo,
2-21-1 Osawa, Mitaka, Tokyo 181-0015}
\altaffiltext{5}{Research Center for Early Universe, 
School of Science, The University of Tokyo, Hongo, 
Bunkyo, Tokyo 113-0033}
\altaffiltext{6}{Max-Planck-Institut f\"ur Radioastronomie,
Auf dem H\"ugel 69, D-53121 Bonn, Germany}
\altaffiltext{7}{Department of Chemistry, University of Virginia,
McCormick Road, PO Box 400319, Charlottesville, VA 22904, USA}
\altaffiltext{8}{Joetsu University of Education, 
Yamayashiki-machi, Joetsu, Niigata 943-8512}


%



\maketitle

\begin{abstract}

Sensitive observations with the Atacama Large Millimeter/submillimeter 
Array (ALMA) allow astronomers
to observe and discuss the detailed distributions of  
molecules with relatively weak intensity in nearby galaxies.  
In particular, we report distributions of several molecular transitions
including shock  and dust related species
($^{13}$CO  $J$ = 1--0, 
C$^{18}$O $J$ = 1--0, 
$^{13}$CN  $N$ = 1--0, 
CS  $J$ = 2--1, 
SO  $J_N$ = 3$_2$--2$_1$, 
HNCO  $J_{Ka,Kc}$ = 5$_{0,5}$--4$_{0,4}$, 
HC$_3$N  $J$ = 11--10, 12--11, 
CH$_3$OH $J_K$ = 2$_K$--1$_K$, and 
CH$_3$CN  $J_K$ = 6$_K$--5$_K$)
in  the nearby Seyfert 2 galaxy NGC 1068.
This is the first paper reporting our study of molecular
distributions in NGC 1068 
with the ALMA early science
program.
The central $\sim$1 arcmin ($\sim$4.3 kpc) 
of this galaxy
was observed in the 100 GHz region (band 3)
covering $\sim$96--100 GHz and $\sim$108--111 GHz
with an angular resolution of $\sim4''\times2''$
(290 pc$\times$140 pc).
These observations were motivated to 
study the effects of an active galactic
nucleus and its surrounding starburst ring on
molecular abundances.
In this article, we present images and 
report a classification of molecular distributions into 
three main categories, defined as follows:  
(1) Molecules concentrated in the circumnuclear disk (CND)
(SO $J_N$ = 3$_2$--2$_1$,  
HC$_3$N $J$ = 11--10, 12--11, and  
CH$_3$CN $J_K$ = 6$_K$--5$_K$),
(2) Molecules distributed both in the CND and the starburst ring
(CS $J$ = 2--1 and
CH$_3$OH $J_K$ = 2$_K$--1$_K$),
(3) Molecules distributed mainly in the starburst ring
($^{13}$CO $J$ = 1--0 and 
C$^{18}$O $J$ = 1--0).
Since most of the molecules such as HC$_3$N 
observed in the CND are easily dissociated
by UV photons and X-rays, our results indicate
that these molecules must be effectively shielded.
In the starburst ring, the distribution of CH$_3$OH 
is similar to those of $^{13}$CO, C$^{18}$O, and CS
on the whole, but the relative intensity of methanol 
at each clumpy region
is not consistent with those of $^{13}$CO, C$^{18}$O,
and CS.
This difference is probably caused by the unique formation and destruction
mechanisms of CH$_3$OH  
in the environment of 
the starburst ring.

\end{abstract}

\section{Introduction}
Recent rapid advances in millimeter/submillimeter (mm/submm) receivers
equipped with  wide-band spectroscopic capabilities,
such as EMIR 
with the WILMA spectrometer
on the IRAM 30m telescope \citep{carter2012},
Z-Spec on the CSO 10.4m telescope 
\citep{bradfo2004}, Redshift-search-receiver on
the LMT 50m \citep{ericks2007},
TZ with the SAM45 spectrometer on the 
NRO 45m telescope \citep{iono2012, nakaji2013},
and SPIRE-FTS on Herschel \citep{griffi2010}, have revolutionalized
our view concerning chemical properties in galaxies. Unbiased spectral line surveys
toward various types of galaxies have been conducted \citep[e.g.,][]{martin2006,
naylor2010, werf2010, snell2011,
costag2011, nakaji2011, martin2011, kamene2011, rangwa2011,
aladro2011, aladro2013}, revealing the richness
and diversity of spectral line features.

These chemical properties have been expected to be powerful astrophysical tools
for the study of galaxies, because activity in the central regions
of galaxies, such as
 a burst of massive star-formation
or an active galactic nucleus (AGN), must have a strong impact
on the chemical and physical properties of the surrounding
interstellar medium (ISM).
For instance, elevated HCN emission with respect to CO and/or HCO$^+$
has often been detected toward AGNs
\citep[e.g.,][]{jackso1993, taccon1994, helfer1995,
kohno1996, kohno2003, krips2007, krips2008, krips2012, izumi2013}, where it is
expected to be the imprint
of  either strong X-ray irradiation/ionization \citep{usero2004,
garcia2010, davies2012}
and/or a high-temperature environment caused by AGN activity
\citep[e.g.,][]{harada2010, izumi2013}. 
Nevertheless,
some controversial observational results
\citep[e.g.,][]{baan2008, snell2011, costag2011, sani2012}
and
theoretical work on the physical and chemical properties of the ISM in external galaxies
\citep[e.g.,][]{meijer2005, meijer2007}, which are
somewhat inconsistent
with the idea of the enhanced HCN emission among AGNs, suggest that our
current understanding
of physical and chemical properties in galaxies is still 
far from complete.

One of the promising directions for study
is to build detailed inventories of spectral lines in the vicinity of AGNs.
By comparing them with the results of  spectral line surveys in the central
regions of starburst galaxies such as NGC 253 
\citep[e.g.,][]{martin2006}
and M82 \citep[e.g.,][]{aladro2011}, we can  
identify the key combinations
of molecules
that differentiate the power sources in galaxies.
For this purpose, the circumnuclear disk 
\citep[CND;][]{usero2004} of NGC 1068
is one of the best targets, 
although NGC 1068 is known
to host intense starburst regions along the inner spiral arms or ring
\citep[e.g.,][]{telesc1988}.
Since the diameter of the starburst ring is 
fairly large ($\sim30''$), the emission
from the CND ($<4''$) can be easily separated spatially, if we employ
mm/submm interferometers.
Note that little if any signature for recent nuclear starburst activity has
been identified in the central region of NGC 1068 
\citep[e.g.,][]{imanis1997, fernan2001, davies2007},
despite the fact that the CND is bright in CO and
other molecular lines \citep[e.g.,][]{schinn2000, tsai2012}.
This makes the CND of NGC 1068 ideal for the study of the AGN imprints,
because nuclear starbursts often
cohabit with AGNs \citep[e.g.,][]{imanis2004}, hampering clear separation
of spectral line features between starbursts and AGNs.

To date, several unbiased line survey works at mm/submm wavelengths
have been reported towards the center of
NGC 1068 
\citep[][]{snell2011, costag2011, kamene2011, spinog2012, aladro2013}
using single dish telescopes, but contamination from starbursts associated
with inner spiral arms/ring could be a problem if we consider the
sizes of the observing
beams 
(14$''$--70$''$). 
On the other hand,
interferometric imaging of the CND give clean measurements of spectral lines,
but the observed lines are limited to major species such as CO, HCN,
HCO$^+$, CS, CN, and SiO
\citep[e.g.,][]{papado1996, taccon1997, kohno2008, garcia2010, krips2011}
due to the limitation in sensitivity of the existing pre-ALMA mm/submm arrays.

In this situation, we proposed to observe several interesting
molecules simultaneously in the 98 and 110 GHz regions sensitively with
ALMA.
These frequency regions are rich in molecules, including 
typical shock/dust related species and the CO isotopologues.
We first obtained the intensities of such lines based on
our line survey project with the 45m telescope \citep{nakaji2011}.
As an additional advantage of this frequency region, 
the primary beams of the ALMA 12m antennas 
($\sim$1 arcmin) cover both the CND
and the starburst ring in one field of view, which
is useful for our purposes.

In this paper, we report a high-resolution imaging study of molecular lines
in the central region ($\sim$1 arcmin) of NGC 1068 
observed with ALMA.
Even in its early science operation phase, ALMA is already powerful enough
to simultaneously observe 10 lines from 9 molecules 
($^{13}$CO $J$ = 1--0, 
C$^{18}$O  $J$ = 1--0, 
$^{13}$CN  $N$ = 1--0, 
CS  $J$ = 2--1, 
SO  $J_N$ = 3$_2$--2$_1$, 
HNCO  $J_{Ka,Kc}$ = 5$_{0,5}$--4$_{0,4}$, 
HC$_3$N  $J$ = 11--10, 12--11, 
CH$_3$OH $J_K$ = 2$_K$--1$_K$, and 
CH$_3$CN $J_K$ = 6$_K$--5$_K$)
within a frequency coverage of $\sim$7.0 GHz at the 
3 mm band, uncovering a wide variety of
molecular line distributions. We describe our observations and data reduction
in Section 2, and present the images and spectra of the observed
molecular lines
in Section 3. 
In Section 4, the molecular distributions will be classified into 
three main
categories: 
(1) Molecules concentrated in the CND, 
(2) Molecules distributed both in the CND
and the starburst ring, and 
(3) Molecules distributed mainly in the
starburst ring.
Some implications of this diversity are also discussed in this section.
Throughout the paper, we assume that the distance of NGC 1068 is 14.4 Mpc
\citep{tully1988, bland1997}; at this distance, $1''$ corresponds to 72 pc.


\section{Observations and data reduction}

The observations were carried out with ALMA in the early 
science program (cycle 0) in January 2012.
The receivers in band 3
(100 GHz region) were used.
The 16 antennas available were in the compact configuration.
The adopted central position of NGC 1068 was 
RA(J2000.0) = \timeform{2h42m40s.798} and  
Dec(J2000.0) = \timeform{-00D00'47".938}.
The systemic velocity employed was 1150 km s$^{-1}$.
The position and the velocity were taken from \cite{schinn2000}.
This position is the radio core at the AGN observed with MERLIN
(The Multi-Element Radio Linked Interferometer Network)
at 5 GHz \citep{muxlow1996}.

As a correlator setup, 
the two spectral windows were placed in the lower sideband 
(LSB: covering $\sim$96--100 GHz),
and the other two were placed in the upper sideband 
(USB: covering $\sim$108--111 GHz).
Each spectral window covered 1875 MHz with 3840 channels
resulting in a frequency resolution of 488 kHz.
Such a setup efficiently covered frequency regions with
rich spectral lines, as mentioned in the Introduction.
The final results were presented with the velocity resolution
of $\sim$19 km s$^{-1}$ (at 100 GHz)
to improve the signal-to-noise ratio. 

The total observational time was about 110 minutes
including calibration and overheads.
The spatial resolution of the observations
was \timeform{4".2}$\times$\timeform{2".4} 
($\sim$300 pc$\times$170 pc)
at the principal axis of 176 deg 
in the spectral window of the lowest frequency. 
The system temperature was $\sim$53--133 K depending on
frequency and antennas.
The achieved noise level (rms) was 
$\sim$1.1--1.7 mJy beam$^{-1}$ depending on the spectral window 
and image region.
The observational parameters are summarized in Table \ref{tab:obs}.

The data were reduced with the reduction 
software CASA (mainly with ver. 3.4).
We used continuum subtracted calibrated data (measurement
set) and image cubes provided by the ALMA Regional Center.
For molecular lines with no provided image cubes
($^{13}$CN, HNCO, and CH$_3$CN), we obtained
images from the measurement set above.
The images of integrated intensity (moment 0 maps)
were made using pixels with all flux values in the
image cubes.


\section{Results}

\subsection{Overview of the images}

\begin{figure}
  \begin{center}
   \FigureFile(170mm,260mm){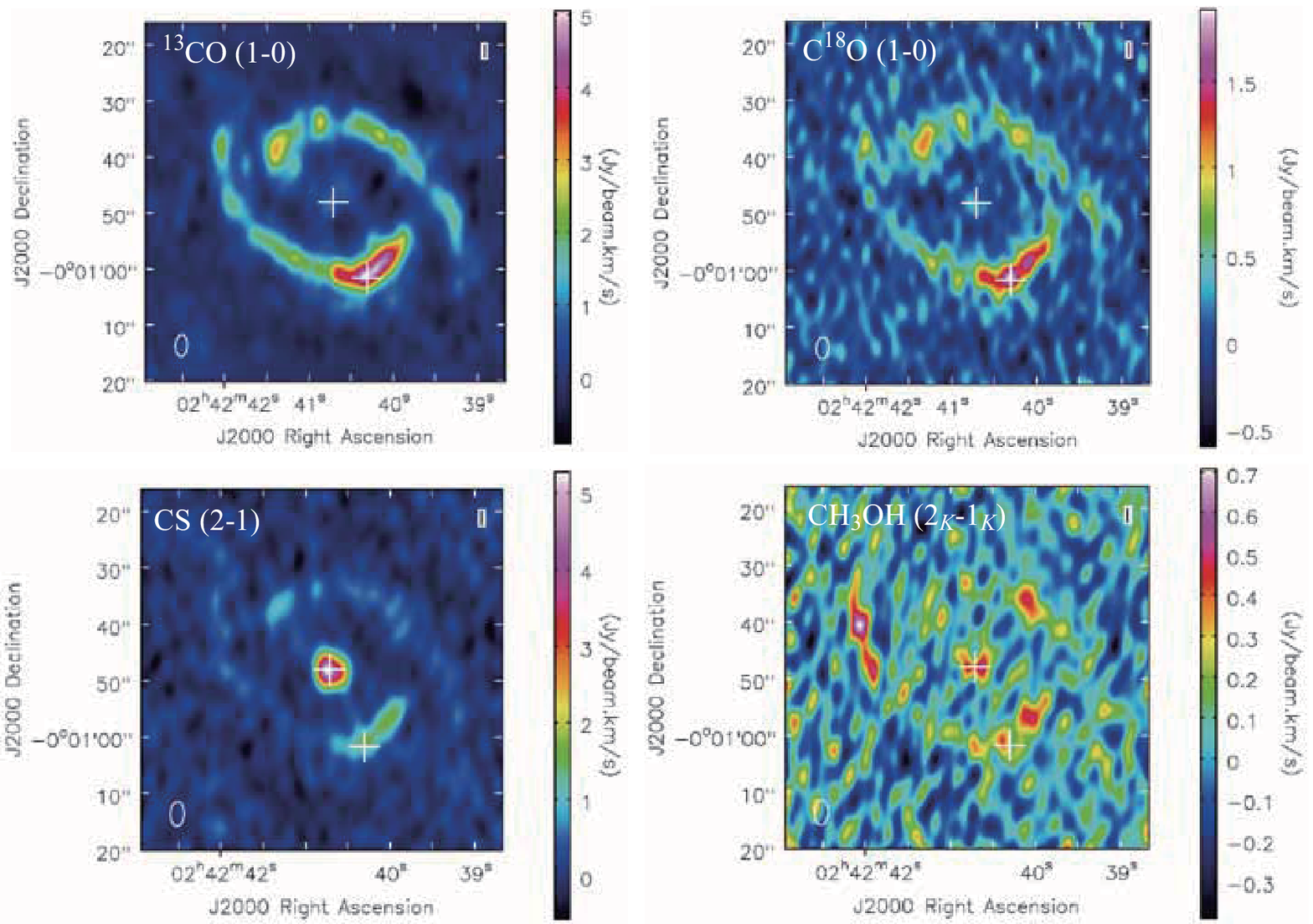}
  \end{center}
  \caption{The images of integrated intensity of 
  $^{13}$CO $J$ = 1--0, C$^{18}$O $J$ = 1--0, CS
  $J$ = 2--1, and CH$_3$OH $J_K$ = 2$_K$--1$_K$.
  The central radio continuum position 
  \citep[RA(J2000.0) = \timeform{2h42m40s.70912} and  
Dec(J2000.0) = \timeform{-00D00'47".9449},][]{gallim2004}
  and the $^{13}$CO 
  $J$ = 3--2 intensity peak at the 
  southwest position in the starburst ring
  (RA(J2000.0) = \timeform{2h42m40s.298} and  
   Dec(J2000.0) = \timeform{-00D01'01".638}, 
   Nakajima et al. in preparation) are 
  indicated with white 
  crosses.
  The beam is shown with an open white ellipse in the bottom-left
  corner in each image.
  The primary beam correction
  is not applied.
   }\label{fig:mom0_1}
\end{figure}

\begin{figure}
  \begin{center}
    \FigureFile(170mm,260mm){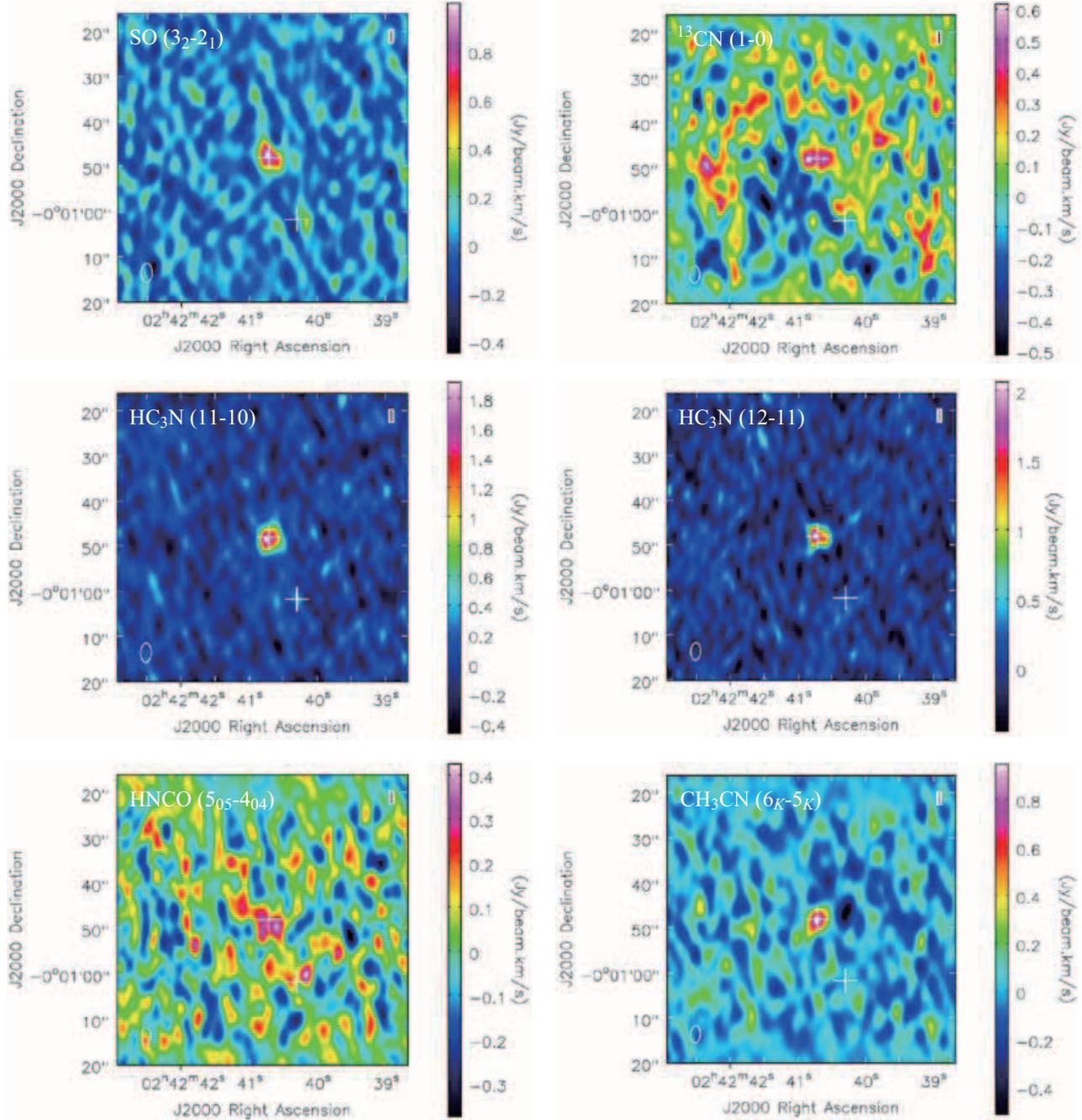}
  \end{center}
  \caption{The images of integrated intensity of 
  SO $J_N$ = 3$_2$--2$_1$,
  $^{13}$CN $N$ = 1--0, 
  HC$_3$N $J$ = 11--10, 12--11, and 
  HNCO $J_{Ka,Kc}$ = 5$_{0,5}$--4$_{0,4}$,
  CH$_3$CN $J_K$ = 6$_K$--5$_K$. 
  The central continuum position and the $^{13}$CO 
  $J$ = 3--2 intensity peak at the 
  southwest position in the starburst ring 
  are indicated with  white 
  crosses (see the caption of figure \ref{fig:mom0_1}).
  The beam is shown with an open
  white ellipse
  in the bottom-left corner in each image.
  The primary beam correction
  is not applied.}\label{fig:mom0_2}
\end{figure}

Thanks to the high sensitivity of ALMA,
all expected lines were detected with a short observational time.
The integrated intensity images are shown in Figures 
\ref{fig:mom0_1} and \ref{fig:mom0_2}.
In Figure \ref{fig:mom0_1} images with a significant
distribution in the 
starburst ring are shown.
These images are  from the 
$^{13}$CO $J$ = 1--0, C$^{18}$O $J$ = 1--0, CS
$J$ = 2--1, and CH$_3$OH $J_K$ = 2$_K$--1$_K$
rotational transitions.
\begin{itemize}
\item $^{13}$CO and C$^{18}$O: The images of these CO isotopic species
show rather weak signals in the CND,  but
show clear  distributions
in the starburst ring.
In particular, a southwest region in the starburst ring shows
the strongest emission.
These overall pictures agree well with the past 
interferometric $^{13}$CO ($J$ = 1--0) images
\citep[e.g.,][]{helfer1995, papado1996, taccon1997},
but the quality of the image is greatly improved with ALMA.  
The clear distributions in the starburst ring are  also 
qualitatively similar to those of the 
$^{12}$CO $J$ = 1--0 and 3--2 transitions
\citep[e.g.,][]{schinn2000, tsai2012},
but the emission in the CND is clearly seen in these
$^{12}$CO images.
The difference in the emission in the CND between $^{12}$CO
and our CO isotopic species indicates that the 
CO $J$ = 1--0 emission in the CND is optically thinner than
that in the starburst ring.
A quantitative analysis will be done with the ALMA 
band 7 data ($\sim$330 GHz region) 
by Nakajima et al. (in preparation) and Taniguchi et al. 
(in preparation).
Next, the total flux ratio of the CO isotopic species 
(C$^{18}$O/$^{13}$CO) was calculated,
and the obtained value is $\sim$0.34. 
This ratio is similar to the corresponding ratio 
of 0.3 from  past interferometric data using the OVRO
(Owens Valley Radio Observatory) mm-array
\citep{papado1996} and the value of 0.28 calculated
from  line survey data with the IRAM 30m telescope
\citep{aladro2013}.  
In addition, we are investigating the properties
of giant molecular associations 
based on 
the $^{13}$CO, C$^{18}$O, CS,
and CH$_3$OH lines
with data cubes of high spectral resolution produced from 
the same data of band 3.
The relation to the star-formation rate (SFR)
is also being investigated.
The results will be published separately (Tosaki et al., 
in preparation).
\item CS: The strong emission is concentrated in the CND, while
 weak emission is seen in the starburst ring.
This is in sharp contrast to the distributions of the CO isotopic species.
Since CS is a typical high-density tracer,  the CS distributed area
should have a relatively high density in the first order approximation.
In the starburst ring, 
the CS emission is relatively strong in the southwestern region.
This pattern is the same as the distributions of the CO isotopic species in the starburst
ring. 
A CS $J$ = 2--1 image  
was also reported by \cite{taccon1997}; these authors
used the IRAM Plateau de Bure interferometer.
Their results also show the central concentration and 
additional clumpy features in the starburst ring.
\item CH$_3$OH: 
Our data yield  the first interferometric image
in NGC 1068.
Methanol is distributed both in the CND and in the starburst
ring with similar intensity.
Although the signal-to-noise ratio is not high enough, 
methanol is 
probably distributed both in the east and west knots 
\citep{schinn2000}
in the CND.
The distribution in the starburst ring is similar to
those of $^{13}$CO, C$^{18}$O, and CS
on the whole, but the relative intensity of methanol 
at each clumpy region
is not as consistent with those of $^{13}$CO, 
C$^{18}$O, and CS.
In particular,
a striking difference can be seen in the eastern region, where
methanol emission is the strongest,
although the intensities of the CO isotopic species 
and CS
are relatively weak.
Methanol is thought to be produced on grain surfaces 
\citep[e.g.,][]{watana2002}
and sublimed into the gas-phase by star-formation activities.
This interesting  
distribution of methanol and its relation to star formation will be discussed later.
\end{itemize}

In Figure \ref{fig:mom0_2}, images 
exhibiting concentrated
distributions
in the CND are shown.
These images involve rotational lines of the following transitions:
$^{13}$CN $N$ = 1--0, SO $J_N$ = 3$_2$--2$_1$, 
HNCO $J_{Ka,Kc}$ = 5$_{0,5}$--4$_{0,4}$, 
HC$_3$N $J$ = 11--10, 12--11, and 
CH$_3$CN $J_K$ = 6$_K$--5$_K$,
and are the first interferometric images for the rotational lines in NGC 1068.
\begin{itemize}
\item $^{13}$CN:
The image was made using two fine structure lines
($J$ = 3/2--1/2 and 1/2--1/2) of the $N$ = 1--0 transition.
Although the signal-to-noise ratio is low, $^{13}$CN may be 
distributed both in the east and west knots in the CND.
It is not clear whether $^{13}$CN is distributed
in the starburst ring, because the clumpy ring-like structure
looks different from those of the CO isotopic species and CH$_3$OH.
Previously, 
the CN $N$ = 2--1 image  
was reported by \cite{garcia2010}, who used the
 IRAM Plateau de Bure interferometer.
Their results show  distributions both 
in the east and west knots in the CND;
the intensity is stronger in the east knot than 
in the west knot.
These facts support our results for $^{13}$CN
in the CND.
Since their field of view is 21", the image does not
cover the starburst ring.
\item SO: 
The image is concentrated in the CND.
\item HNCO: Although the signal-to-noise ratio is low, HNCO may be 
distributed both in the east and west knots in the CND, and
is distributed in the starburst ring, though it is not
clearly seen in the image.
As shown later, the distribution of HNCO in the starburst ring
is confirmed from the detection of its spectral line
in the southwest point of the starburst ring.
\item HC$_3$N: The image is concentrated in the CND as seen from
 two rotational emission lines.
\item CH$_3$CN: The image is concentrated in the CND.
\end{itemize}
\subsection{Spectra at the circumnuclear disk and the 
starburst ring (southwest)}
\begin{figure}
  \begin{center}
    \FigureFile(125mm,200mm){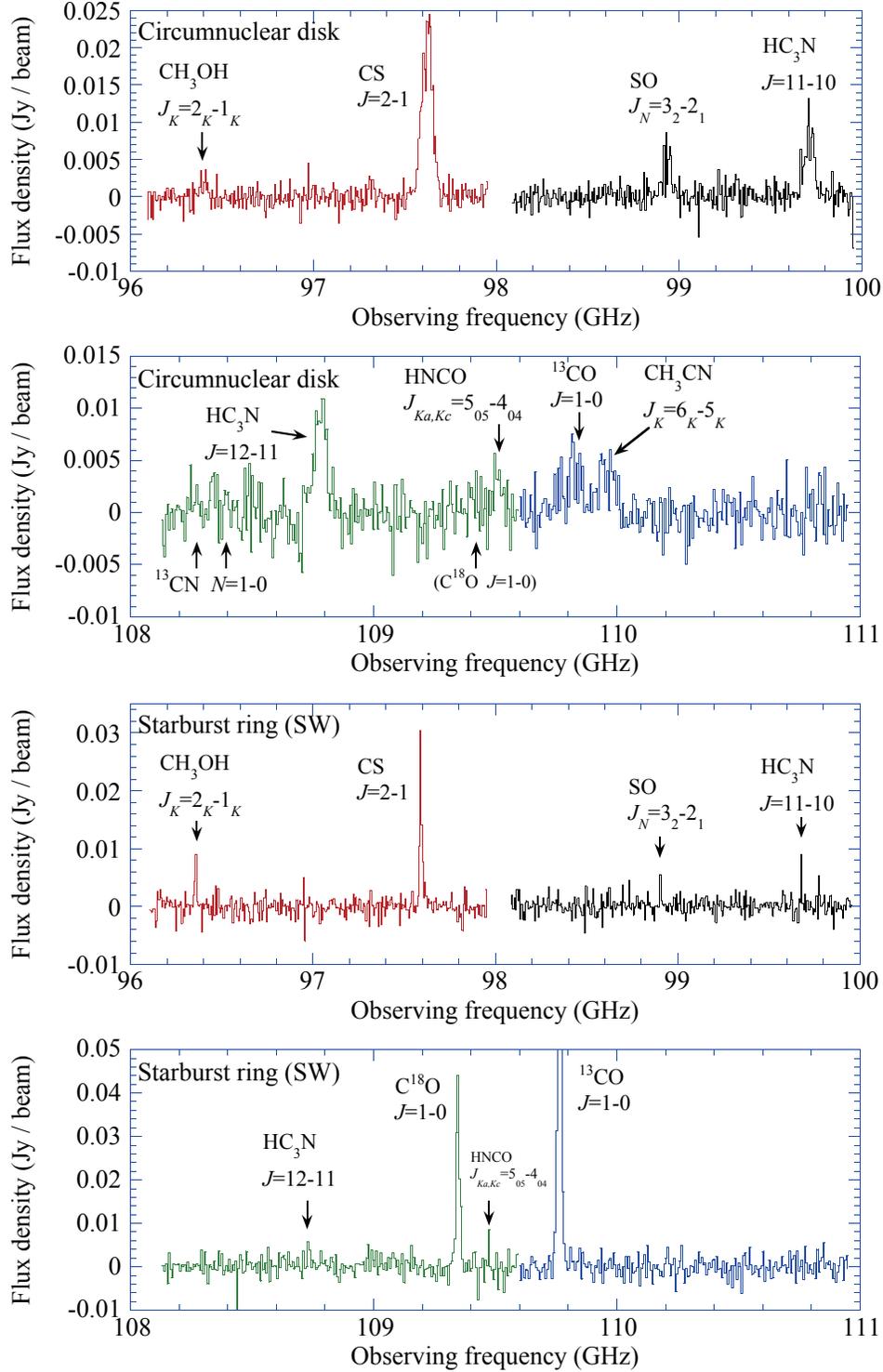}
  \end{center}
  \caption{ 
  The spectra are shown
  at the central continuum position and the 
  $^{13}$CO $J$ = 3--2  intensity peak at the 
  southwest position in the starburst ring 
  (see the caption of figure \ref{fig:mom0_1}).
  The primary beam correction
  is applied.
  Different colors of the spectra indicate different
  spectral windows (spw0, 1, 2, and 3).
  The frequency is shown as topocentric value, which is a
  default reference frame of ALMA.
  It is necessary to shift the frequency corresponding to
  $V_{\rm LSR}$ = 1150 km s$^{-1}$ to obtain 
  approximate rest frequency.
  }\label{fig:4spectra}
\end{figure}

Spectra were obtained from cleaned images at the following two positions:
\\
(1) The central radio continuum peak (AGN) in the CND
\citep[RA(J2000.0) = \timeform{2h42m40s.70912} and  
Dec(J2000.0) = \timeform{-00D00'47".9449},][]{gallim2004}
\\
(2) The $^{13}$CO $J$ = 3--2 intensity peak at the 
southwest position in the starburst ring
(RA(J2000.0) = \timeform{2h42m40s.298} and  
   Dec(J2000.0) = \timeform{-00D01'01".638}, 
   Nakajima et al. in preparation). 
\\
Before obtaining the spectra, 
the images in the two spectral windows
in the USB were convolved with the beam of the 
spectral window in the 
lowest frequency in the LSB, 
and then, 
attenuation due to the 
primary beam 
pattern of the ALMA 12m antennas was 
corrected. 
The velocity resolution is $\sim$19.0 km s$^{-1}$
at 100 GHz.
The obtained spectra are shown in Figure
\ref{fig:4spectra}.

In the CND, the detected lines are broad, with a width of about 
200 km s$^{-1}$ FWHM.
C$^{18}$O is not clearly detected, and  
$^{13}$CN is marginally detected in the CND.
On the other hand, the detected lines 
in the southwest position of the starburst ring are narrow 
with a width of  about 20--45 km s$^{-1}$ FWHM. 
Detections of SO $J_N$ = 3$_2$--2$_1$, 
HC$_3$N $J$ = 11--10, 12--11, and
HNCO $J_{Ka,Kc}$ = 5$_{0,5}$--4$_{0,4}$
clearly indicate the existence of such
molecules there, which is not unambiguously
determined by the images in figure \ref{fig:mom0_2}.
The CH$_3$CN $J_K$ = 6$_K$--5$_K$ transition is not
detected.
The detected lines were Gaussian fitted, and the
 line parameters obtained are summarized in 
Table \ref{tab:param}.


\subsection{Total flux and the flux at the circumnuclear disk}
As presented in Figures \ref{fig:mom0_1} and \ref{fig:mom0_2},
the distributions of molecules show wide diversity. 
In order to obtain quantitative information of the distributions,
we extracted both the total flux and the flux in the CND
for each molecule.  
The derived flux ratio
(CND/Total) 
was 
then used to determine a general trend.

The total flux was obtained from the images 
within a circle of \timeform{55"} diameter,
which covers both the CND and the starburst ring.
The flux in the CND is obtained within a circle of 
\timeform{10"} diameter.
Both of the circles are centered on the 
radio continuum peak \citep{gallim2004}.
The results are listed in Table  \ref{tab:flux}.
The error for each flux is obtained from the 
standard deviation of the pixel values of the image,
but the systematic deviation is not included.
For images with low signal-to-noise ratio, 
it is difficult to obtain a reliable total flux from the 
relatively wide area of the \timeform{55"} 
diameter, because the effect of  
systematic deviation in the image can be significant. 
In such a case, the obtained values were not presented 
in this table.
We also note that the accuracy of the $^{13}$CO flux 
in the CND may be limited by the dynamic range of 
the data (Taniguchi et al., in preparation), 
because the emission from the CND is surrounded by 
much stronger sources of emission along the starburst ring. 

\begin{figure}
  \begin{center}
    \FigureFile(90mm,90mm){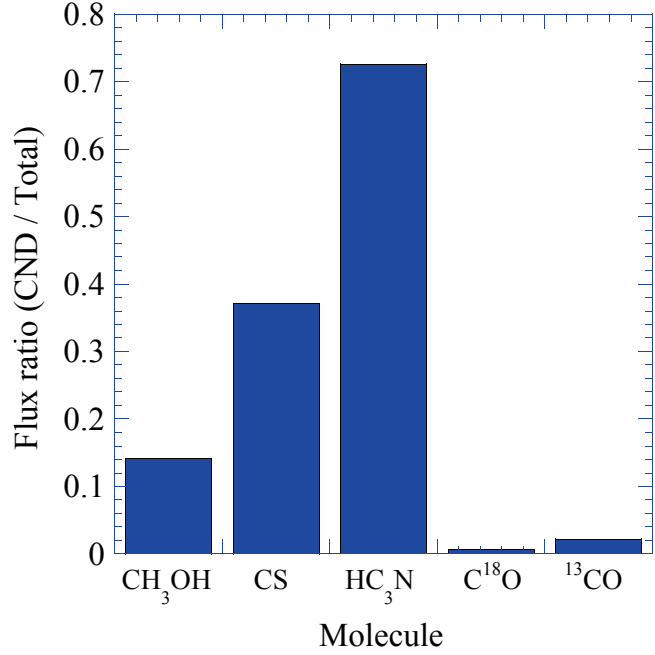}
  \end{center}
  \caption{ 
  Flux ratio (circumnuclear disk (CND) / total) is presented
  for each molecule.
  For details, see the text (section 3.3).
  
  }\label{fig:flux_ratio}
\end{figure}

The value of the flux ratio, which is presented 
in Table  \ref{tab:flux},  ranges widely from
0.006 to more than 0.7.
By comparing the ratio and the image,
we can approximately establish the correspondence between
the ratio and the distribution 
as follows:
\\
(1) 0.7--1.0 for molecules concentrated in the CND,
\\
(2) 0.1--0.4 for molecules distributed both in the CND and the starburst ring,
\\
(3) 0.0--0.02 for molecules distributed mainly in the starburst ring.
\\
The flux ratio is also presented in 
Figure \ref{fig:flux_ratio}.

\subsection{Comparison of flux with single dish telescopes}
We compared the flux obtained from ALMA with those 
from
single dish telescopes (NRO 45m and IRAM 30m) 
to estimate the recovery of the 
flux with the interferometer.
First, the flux of $^{13}$CO was compared.
The image obtained with ALMA, with the 
primary beam corrected,
was convolved with the 
45m beam of \timeform{16"}, and
the flux obtained was converted to a
brightness temperature of $\sim$8.1 K km s$^{-1}$.
The corresponding value obtained with the 45m telescope
is $\sim$8.9 K km s$^{-1}$, which was observed in the 
recent line survey project
(Nakajima et al. and 
Takano et al., in preparation).
Therefore, the recovered flux is about 92 \%.
The same comparison was carried out by convolving the ALMA data
with the 30m beam of \timeform{22"}, and
the flux obtained was once again converted to 
brightness temperature, this time $\sim$10.1 K km s$^{-1}$.
The corresponding value obtained with the 30m telescope
is $\sim$12.8 K km s$^{-1}$ \citep{aladro2013}, so that the recovered flux is about 79 \%.
The 30m telescope can observe widely distributed gas
more efficiently than the 45m telescope
due to the larger beam size, 
but ALMA
is less sensitive to such gas.
The difference in the values of the recovery  
may
include such an effect.

Second, the flux of CS, which is more compactly distributed
than $^{13}$CO, was compared.
The image obtained with ALMA, with  
primary beam corrected,
was convolved with the 
45m beam of \timeform{17"}, and
the flux obtained was converted to a
brightness temperature of $\sim$5.9 K km s$^{-1}$.
The corresponding value obtained with the 45m telescope
is $\sim$8.0 K km s$^{-1}$
(Nakajima et al. and 
Takano et al., in preparation).
Therefore, the recovered flux is about 74 \%.
The comparison with the CS data of the 30m telescope
was not done, because the line is blended with 
a line from the other sideband \citep{aladro2013}. 
These results of $^{13}$CO and CS 
indicate that 
ALMA observes a
significant amount of gas in the central region of
NGC 1068.
 

\section{Discussion}

\subsection{Classification of molecular distributions}
As presented in Figures \ref{fig:mom0_1} and \ref{fig:mom0_2},
the molecules exhibit a wide variety of distributions,
which are reflections of abundance and excitation.
Such distributions contain important information 
to study the effects on molecules caused by
 AGN activity and  starburst conditions.
As already mentioned in the previous sections, we can classify
molecular distributions into three broad categories:
\\
(1) Molecules concentrated in the CND, 
\\
(2) Molecules distributed both in the CND
and the starburst ring,
\\ 
(3) Molecules distributed mainly in the
starburst ring.
\\
Based in addition on the spectra shown in Figure
\ref{fig:4spectra} and on the flux ratio (CND/Total)
in Table \ref{tab:flux}, each molecule was 
further classified as listed in Table \ref{tab:class},
where
the above categories (1) and (2) were subdivided
as follows:
\begin{itemize}
\item Molecules distributed in two knots in the CND,
\item Molecules distributed in the center of the CND.
\end{itemize}
Higher spatial resolution and better signal-to-noise 
ratio are desirable for this more precise classification,
but indications of such differences can be seen in the
present images.
In Table \ref{tab:class}, 
the refined classification of 
HCN ($J$ = 1--0), 
HCO$^+$ ($J$ = 1--0), 
and SiO ($J$ = 2--1) is 
also included based on their images in the literature
\citep[e.g.,][]{jackso1993, kohno2008, garcia2010}.
Distributions of 
$^{13}$CO, C$^{18}$O, CN, and CS  
obtained from our ALMA band 7 data
(Nakajima et al. in preparation) are also included.

\subsection{Implications for molecular formation and 
destruction mechanisms}

We discuss the implications of specific molecular distributions below
using a knowledge of the relevant reactions and the results of
model calculations.
Further discussion will be undertaken after we obtain
quantitative information such as 
the abundances in Nakajima et al. (in preparation).
Here we mainly use the results of recent models 
by \cite{harada2010, harada2013} and \cite{aladro2013}. 
\cite{harada2010} 
reported a gas-phase time-dependent reaction model including reactions with significant activation energies
for high-temperature chemistry.
Thus, this model can be used for high-temperature sources up to
$\sim$800 K.
\cite{harada2013} applied the results of \cite{harada2010} to 
the axisymmetric accretion disk around AGNs such as that in NGC 1068.
The effects of X-rays and cosmic rays are included.
\cite{aladro2013} reported a  time-dependent reaction model 
including gas-phase and grain surface reactions.
The effects of UV photons and cosmic rays are included. 
A high cosmic ray ionization rate was used to simulate the environment
in an X-ray dissociation region (XDR).

\subsubsection{HC$_3$N}
Because HC$_3$N is concentrated in the CND, its behavior
 cannot be explained if we consider the 
entire CND as an XDR, 
since HC$_3$N is 
easily dissociated by cosmic rays and UV photons 
\citep{aladro2013, harada2013}.
On the other hand, the abundance of  HC$_3$N
increases at hot core regions 
and high-temperature (non-dissociative shocked) 
regions \citep{casell1993, harada2010}.
The model by \cite{harada2013} shows that 
there is a 
high-temperature 
midplane where X-rays cannot penetrate into  the CND
in addition to a different layer that can be characterized as an XDR. 
Complex molecules are predicted to be abundant in the high temperature zone.
Our observations of HC$_{3}$N  confirm that there is a 
large amount of gas which is shielded from X-rays, 
and the local column densities of the CND should be very high. 
This fact yields information on the  
structure of the CND, which is  quite interesting and relevant to an
understanding of the physical and chemical environment
in the central region.
Further ALMA observations with higher angular resolution
will reveal the structure.

\subsubsection{CH$_3$OH}
Methanol has a significant concentration in the CND as mentioned before.
Just like HC$_3$N, methanol is 
easily dissociated by cosmic rays and UV photons \citep{aladro2013}.
Thus, the concentration of methanol in the CND also requires shielding.
On the other hand,
the 
inconsistent
distribution of methanol in the 
starburst ring is very striking.
The formation of methanol on icy dust grains is well known and it is not surprising that methanol should be
abundant in active star-forming regions in the starburst ring.

To look at the matter further, we 
studied the relation of methanol intensity to 
the star-formation rate 
\citep[SFR,][]{tsai2012} in those regions with
relatively strong
methanol emission.
We found the averaged SFR in regions with relatively
strong methanol emission 
(roughly at regions R6, R9, 
R16, and R20--21 in \cite{tsai2012}) to be 
1.40$\pm$0.54
$\MO$ yr$^{-1}$ kpc$^{-2}$, whereas the average of 
all regions is 1.26$\pm$0.46
$\MO$ yr$^{-1}$ kpc$^{-2}$.
The averaged value of regions with relatively
strong methanol emission
is only slightly larger than that of  all the regions.
Moreover, the values at 
 regions with relatively
strong methanol emission
have a large scatter from 0.97 to 2.26 
$\MO$ yr$^{-1}$ kpc$^{-2}$ \citep{tsai2012}
as indicated in the large 
standard deviation. 
Thus,
we could not find a strong correlation between
 methanol intensity and the SFR.
This result may be related to our spatial resolution and
its coupling to the sizes of the regions above, 
or this may be due to the formation and destruction
mechanisms in the environment of the starburst ring.
A more detailed 
analysis will be published separately
(Tosaki et al. in preparation).

The inconsistent distribution of 
CH$_3$OH in the starburst ring when compared with 
$^{13}$CO and/or C$^{18}$O, discussed here,
was also reported in IC 342 \citep{meier2005}
and Maffei 2 \citep{meier2012}.
In addition, CH$_3$OH is relatively weak in M82
\citep[e.g.,][]{aladro2011},
which is in a late stage of starburst activity.
These  presumably related results should be very helpful in
interpreting the inconsistency in the starburst ring 
in NGC 1068, suggesting that methanol needs specific favorable
conditions to form on grains and to 
subsequently sublime into the gas phase.

\subsubsection{SO, HNCO, and CH$_3$CN}
Because 
they are concentrated in the CND, the distributions of these molecules
 clearly indicate that the CND is a good 
environment to maintain their abundances.
SO and CH$_3$CN are easily dissociated by UV photons, but
the abundance of SO is enhanced by cosmic rays \citep{aladro2013}.
Thus, the XDR environment seems to be a favorable place for SO, but
the concentration of CH$_3$CN also suggests shielded
gas in the CND. 

Similar to other molecules, HNCO can also be dissociated by UV
photons, and it needs to be shielded.
Although there is a gas-phase production route for this molecule, 
its abundance is known to increase when HNCO on dust surfaces sublimes
\citep{quan2010}.

The above mentioned molecules 
except CH$_3$CN 
(SO, HNCO, HC$_3$N, and CH$_3$OH)  are also 
detected in the southwest starburst ring.
These facts do not contradict the detections of 
these molecules in other starburst galaxies such as NGC 253
and M82
\citep[e.g.,][]{martin2006, aladro2011}.

\subsubsection{Two kinds of distributions in the CND}
The molecules in the CND show two
kinds of distributions as mentioned previously in  
subsection 4.1.
Here we briefly discuss possible reasons for such a difference.
Let us start with the case of SiO,
which
 is distributed in the center, as observed by \cite{garcia2010}, who also
 reported an enhanced SiO abundance and discussed its likely shock origin. 
Probably SiO is produced through the sputtering 
of Si-bearing material in grains
\citep{field1997, casell1997,schilk1997}, and as a result
SiO traces strong shocks in the central region.
On the other hand, 
HNCO and CH$_3$OH are tracing relatively weak shocks,
which result in the sublimation of HNCO and CH$_3$OH from the icy mantles of dust grains
\citep[e.g.,][]{rodrig2010, bachil1995}.
Therefore, it seems that  molecules produced in more energetic
regions tend to exist closer to the center, 
while those produced in less energetic regions exhibit a bipolar pattern in the knots.

As shown in this study,
the sensitive ALMA observations are very powerful probes of the
distributions of molecules 
in gas-rich nearby galaxies such as NGC 1068
even in the early operation phase of ALMA.
Further observations of many other molecules will bring us 
additional valuable information to understand chemical
processes in such environments.

\section{Summary}
We have observed the Seyfert 2 galaxy NGC 1068 
in its central 1 arcmin at the 100 GHz frequency region during 
the ALMA early science program; our observations included
 both the 
CND and the starburst ring.

We observed the rotational transitions 
$^{13}$CO $J$ = 1--0, 
C$^{18}$O $J$ = 1--0, 
$^{13}$CN $N$ = 1--0, 
CS $J$ = 2--1, 
SO $J_N$ = 3$_2$--2$_1$, 
HNCO $J_{Ka,Kc}$ = 5$_{0,5}$--4$_{0,4}$, 
HC$_3$N $J$ = 11--10, 12--11, 
CH$_3$OH $J_K$ = 2$_K$--1$_K$, and 
CH$_3$CN $J_K$ = 6$_K$--5$_K$.
The molecular transitions show
a wide variety of spatial distributions, which can be classified
into three main categories: those molecular 
transitions concentrated in the CND, those 
distributed both in the CND and the starburst ring, 
and those mainly in the starburst ring.  
The distributions concentrated in the CND include $^{13}$CN, 
SO, 
HNCO, 
HC$_3$N, and
CH$_3$CN; those distributed in the CND and starburst ring include  
CS and
CH$_3$OH, 
while those mainly in the starburst ring include
$^{13}$CO and C$^{18}$O.
The first two categories  were further
subdivided into distributions in the two knots or 
in the center of the CND.

 Molecules concentrated in the CND 
are easily dissociated
by cosmic rays, X-rays, and/or UV photons. 
These facts indicate that 
there is a 
large amount of gas in the CND, which is shielded 
especially from X-rays.
This information constrains the  
structure of the CND.

 The distribution of methanol in the starburst ring
shows an inconsistent relative intensity with respect
to those of $^{13}$CO, C$^{18}$O, and CS.
In addition, the intensity of methanol does not seem to
correlate with the star formation efficiency, although
methanol is thought to be formed on grain surfaces and sublimed into the
gas-phase by external heating.


--------------------

This paper makes use of the  ALMA data 
ADS/JAO.ALMA\#2011.0.00061.S. 
ALMA is a partnership of ESO (representing its 
member states), NSF (USA) and NINS (Japan), 
together with NRC (Canada) and 
NSC and ASIAA (Taiwan), in cooperation with 
the Republic of Chile. 
The Joint ALMA Observatory is operated by 
ESO, AUI/NRAO and NAOJ.
We thank the support of the East Asian ALMA Regional Center,
in particular, A. Kawamura, for the support.
S. T. thanks Y. Shimajiri, M. Oya, and S. Takahashi for 
the support of the analysis with CASA.   
E. H. wishes to acknowledge the support of the 
National Science Foundation for his astrochemistry program. 
He also acknowledges support from the NASA Exobiology 
and Evolutionary Biology program through a subcontract 
from Rensselaer Polytechnic Institute.


\clearpage

\bibliographystyle{aa}
\bibliography{takano-gal}


\newpage

\begin{table}
  \caption{Observational parameters (ALMA Band 3, cycle 0)}\label{tab:obs}
  \begin{center}
    \begin{tabular}{ll}
      \hline
      Parameter & Value  \\
      \hline
      Date              &    January 9 and 10, 2012  \\
      No. of antennas     &  16  \\      
      Configuration     &   compact  \\     
      Phase center \citep{schinn2000}     &   RA(J2000.0) = \timeform{2h42m40s.798}   \\
                       &   Dec(J2000.0) = \timeform{-00D00'47".938} \\
      Bandpass calibrator              &  J0423-013     \\
      Flux calibrator                  &  Callisto     \\
      Phase calibrator                 & J0339-017      \\
%
      Central freq. (GHz) and beam size   & 97.38 (LSB, spw0), \timeform{4.2"}$\times\timeform{2.4"}$ 176 deg           \\
      \ \ with its principal axis  & 99.3875 (LSB, spw1), \timeform{4.2"}$\times$\timeform{2.2"} 178 deg           \\
      \ \ of each spectral window  &  109.4375 (USB, spw2), \timeform{3.8"}$\times$\timeform{2.2"} 178 deg        \\
                               &  110.4375 (USB, spw3), \timeform{3.9"}$\times$\timeform{2.1"} 177 deg        \\   
      Frequency resolution (kHz) &  488        \\   
      Velocity resolution (km s$^{-1}$)      &   $\sim$19.0  at 100 GHz (13 channel binning)    \\
      Rms noise (mJy beam$^{-1}$)             &   $\sim$1.1--1.7    \\                                                                      
      \hline     
    \end{tabular}
  \end{center}
\end{table}

\tiny
\begin{landscape}
\begin{longtable}{lllllllllllll}
\caption{Line parameters of the detected lines} \label{tab:param}\\
\hline              
Frequency$^*$ & Molecule & Transition 
& \multicolumn{4}{c}{Circumnuclear disk} &
& \multicolumn{4}{c}{Starburst ring (SW)$^{\dagger}$} 
& Comment  \\
\cline{4-7}
\cline{9-12}
 &  &  
& Flux density & $V_{\rm LSR}$ & FWHM & $\int {\rm flux} dv$  &
& Flux density & $V_{\rm LSR}$ & FWHM & $\int {\rm flux} dv$   
&   \\

 (MHz)        &       &           
& (mJy        & (km s$^{-1}$) & (km s$^{-1}$) & 
(Jy beam$^{-1}$  & 
& (mJy  & (km s$^{-1}$) & (km s$^{-1}$) & 
(Jy beam$^{-1}$  &           
             \\ 

               &       &           
& beam$^{-1}$) &       &      & km s$^{-1}$)    &
& beam$^{-1}$) &       &      & km s$^{-1}$)    &             \\ 
  \hline
\endhead
  \hline
\endfoot
  \hline
  \multicolumn{13}{l}{\hbox to 0pt{\parbox{230mm}{\tiny
    \footnotemark[$^*$] Frequency measured in laboratory \citep{lovas1992}: 
    $J_{Ka,Kc}$=2$_{0,2}$--1$_{0,1} A+$
    for CH$_3$OH and
    $J_K$=6$_0$--5$_0$ for CH$_3$CN.  
    \par\noindent
    \footnotemark[$^{\dagger}$] Values at the peak channel are listed for the
    HC$_3$N $J$=12--11 and HNCO $J_{Ka,Kc}$=5$_{0,5}$--4$_{0,4}$
    transitions, since Gaussian
    fitting could not be applied.  
    For these transitions, the errors are calculated from the rms noise
    in the spectra.  
    \par\noindent
    \footnotemark[$^{\ddagger}$]  
    Transitions contributing to this 2$_K$--1$_K$ group are
    $J_{Ka,Kc}$=2$_{-1,2}$--1$_{-1,1} E$, 2$_{0,2}$--1$_{0,1} A+$,
    2$_{0,2}$--1$_{0,1} E$, and 2$_{1,1}$--1$_{1,0} E$.  
    \par\noindent
    }}}
\endlastfoot
  \hline

96741.42 & CH$_3$OH & $J_K$=2$_K$--1$_K$$^{\ddagger}$
& 2.3$\pm$0.6 & 1057$\pm$20 & 148$\pm$51 & 0.4$\pm$0.1 & 
& 9.8$\pm$1.8 & 1190$\pm$3 & 29$\pm$7 & 0.31$\pm$0.06 & 
  \\

97980.968 & CS & $J$=2--1
& 22.0$\pm$0.9 & 1092$\pm$4 & 209$\pm$11 & 4.9$\pm$0.2 & 
& 29$\pm$2 & 1193$\pm$1 & 37$\pm$2 & 1.13$\pm$0.07 &   \\


99299.879 & SO & $J_N$=3$_2$--2$_1$
& 5.1$\pm$0.8 & 1090$\pm$13 & 166$\pm$33 & 0.9$\pm$0.2 & 
& 6.5$\pm$3.5 & 1188$\pm$6 & 20$\pm$12 & 0.14$\pm$0.08 
&  \\

100076.389 & HC$_3$N & $J$=11--10
& 9.1$\pm$0.9 & 1083$\pm$8 & 178$\pm$21 & 1.7$\pm$0.2                 &
& 9.0$\pm$1.6 & 1190$\pm$6 & 19$\pm$6 & 0.19$\pm$0.05 
&  \\
        
108657.646 & $^{13}$CN & $N$=1--0, $J$=1/2--1/2 
& --- & --- & --- & ---                 &
& --- & --- & --- & $<$0.08 (1$\sigma$)
&  low SN (CND, SB ring) \\

108780.201 & $^{13}$CN & $N$=1--0, $J$=3/2--1/2 
& --- & --- & --- & ---                 &
& --- & --- & --- & $<$0.08 (1$\sigma$)
&  low SN (CND, SB ring) \\

109173.634 & HC$_3$N & $J$=12--11 
& 10.6$\pm$0.9 & 1079$\pm$7  & 171$\pm$17 & 1.9$\pm$0.2 & 
& 5.8$\pm$2.0  & 1216  & $\sim$30  &  0.17$\pm$0.05 
&  low S\/N (SB ring)  \\

109782.160 & C$^{18}$O & $J$=1--0 
& --- & --- & --- & $<$0.12 (1$\sigma$) & 
& 42$\pm$2 & 1191$\pm$1 & 45$\pm$3 & 2.0$\pm$0.1 
&  \\

109905.753 & HNCO & $J_{Ka,Kc}$=5$_{0,5}$--4$_{0,4}$ 
& 3.6$\pm$1.1 & 1078$\pm$15 & 95$\pm$37 & 0.4$\pm$0.1                 &
& 8.5$\pm$2.4   &  1179 &  17 & 0.15$\pm$0.04  
& low S\/N (CND, SB ring)  \\

110201.353 & $^{13}$CO & $J$=1--0 
& 4.0$\pm$0.8 & 1040$\pm$24    & 254$\pm$61   & 1.1$\pm$0.2                &
& 131$\pm$2   & 1191.2$\pm$0.4 & 44.5$\pm$0.9 & 6.2$\pm$0.1 
&  \\

110383.522 & CH$_3$CN & $J_K$=6$_K$--5$_K$ 
& 4.5$\pm$1.0 & 1169$\pm$19 & 237$\pm$70 & 1.1$\pm$0.3                 &
& ---         & ---         & ---        & $<$0.10 (1$\sigma$)           &   \\

\hline
\end{longtable}

\end{landscape}

\begin{table}
  \caption{Flux in the circumnuclear disk (CND) and the total flux of each molecule}\label{tab:flux}
  \begin{center}
    \begin{tabular}{lllll}
      \hline
      Molecule & Flux in the CND\footnotemark[$^*$] & Total flux\footnotemark[$^\dagger$] & 
      Ratio (CND/Total) & Main distribution \\
               & (Jy km s$^{-1}$) & (Jy km s$^{-1}$) &   \\
      \hline
CH$_3$OH           & 1.05$\pm$0.03 & 7.4$\pm$0.2     &  0.141$\pm$0.005 & CND \& starburst ring\\ 
CS                 & 10.4$\pm$0.3  & 28.0$\pm$0.5    & 0.37$\pm$0.01   & CND \& starburst ring \\  
SO                 & 1.78$\pm$0.05 & ---\footnotemark[$^\ddagger$]    & ---     & CND  \\  
HC$_3$N $J$=11--10 & 3.7$\pm$0.1   & ---\footnotemark[$^\ddagger$]    & ---     & CND  \\
HC$_3$N $J$=12--11 & 3.3$\pm$0.1   & 4.6$\pm$0.3        & 0.73$\pm$0.06   & CND  \\
$^{13}$CN          & 1.55$\pm$0.05 & ---\footnotemark[$^\ddagger$]      & ---   & (at least) CND  \\ 
C$^{18}$O          & 0.34$\pm$0.03 & 48.0$\pm$0.5    & 0.0063$\pm$0.0006 & starburst ring\\  
HNCO               & 1.27$\pm$0.03 & ---\footnotemark[$^\ddagger$]   & ---  & (at least) CND  \\
$^{13}$CO          & 3.07$\pm$0.07 & 143$\pm$1       & 0.0217$\pm$0.0005 & starburst ring\\
CH$_3$CN           & 1.61$\pm$0.06 & ---\footnotemark[$^\ddagger$]   & ---     & CND \\                                                                          
      \hline     
      \multicolumn{2}{l}{\hbox to 0pt{\parbox{180mm}{\footnotesize
    \footnotemark[$^*$] Flux within the circle with the diameter of 10".
    \par\noindent
    \footnotemark[$^\dagger$] Flux within the circle with the diameter of 55".
    \par\noindent
    \footnotemark[$^\ddagger$] In the case of low signal-to-noise ratio, it is difficult to 
    obtain a reliable flux of the relatively wide area of 55" diameter.
       \par\noindent
    For details, see the text (section 3.3).
    }}}
    \end{tabular}
  \end{center}
\end{table}
\begin{table}
  \caption{Classification of molecular distributions in 
  the circumnuclear disk (CND) and the starburst ring}\label{tab:class}
  \begin{center}
    \begin{tabular}{l}
      \hline
       Category and Molecule   \\
      \hline
      1. Molecules concentrated in the CND            \\
         \ \ \ \ 1--1. Distributed both in the east and west knots   \\ 
         \ \ \ \ \ \ \ \ $^{13}$CN ($N$ = 1--0)\footnotemark[$^*$], 
         HNCO ($J_{Ka,Kc}$ = 5$_{0,5}$--4$_{0,4}$)\footnotemark[$^*$]\footnotemark[$^\dagger$],
         CN ($N$ = 3--2)\footnotemark[$^\|$], 
         CS ($J$ = 7--6)\footnotemark[$^\|$] 
         \\ 
         \ \ \ \ 1--2. Distributed in the center                 \\
         \ \ \ \ \ \ \ \ SO ($J_N$ = 3$_2$--2$_1$)\footnotemark[$^\dagger$],
         SiO ($J$ = 2--1)\footnotemark[$^\ddagger$],
         HC$_3$N ($J$ = 11--10, 12--11)\footnotemark[$^\dagger$], 
         CH$_3$CN ($J_K$ = 6$_K$--5$_K$)                    \\
      2. Molecules distributed both in the CND
                         and the starburst ring         \\
         \ \ \ \ 2--1. Distributed both in the east and west knots in the CND   \\
         \ \ \ \ \ \ \ \   CH$_3$OH ($J_K$ = 2$_K$--1$_K$),
         $^{13}$CO ($J$ = 3--2)\footnotemark[$^\|$]
                                  \\ 
         \ \ \ \ 2--2. Distributed in the center in the CND                  \\
         \ \ \ \ \ \ \ \       CS ($J$ = 2--1),                
         HCN ($J$ = 1--0)\footnotemark[$^\S$], 
         HCO$^+$ ($J$ = 1--0)\footnotemark[$^\S$]                 \\
      3. Molecules distributed mainly in the
                         starburst ring.               \\  
         \ \ \ \ \ \ \ \       $^{13}$CO ($J$ = 1--0), 
                C$^{18}$O ($J$ = 1--0, 3--2\footnotemark[$^\|$])                    \\                                                                
      \hline     
      \multicolumn{1}{l}{\hbox to 0pt{\parbox{180mm}{\footnotesize
    \footnotemark[$^*$] Note that the signal-to-noise ratio is not high.
    \par\noindent
    \footnotemark[$^\dagger$] Relatively small portion of the flux is detected 
    in the southwestern part of the starburst ring 
    (see figure \ref{fig:4spectra}).
    \par\noindent
    \footnotemark[$^\ddagger$] Based on \cite{garcia2010}
    \par\noindent
    \footnotemark[$^\S$] Based on \cite{jackso1993}, \cite{taccon1994}, 
    and \cite{kohno2008}
    \par\noindent
    \footnotemark[$^\|$] See Nakajima et al. in preparation.
    }}}
    \end{tabular}
  \end{center}
\end{table}

\end{document}